\documentclass[12pt]{article}

\usepackage[utf8]{inputenc}
\usepackage{amsmath,amsfonts,amsthm}
\usepackage[a4paper,total={16cm,26cm}]{geometry}
\usepackage[small]{titlesec}
\usepackage[dvipsnames]{xcolor}
\usepackage{mathrsfs} % \mathscr{} 
\usepackage{hyperref}
\usepackage{tikz}
\usepackage{setspace}
\usepackage{tcolorbox}
\usepackage{algorithm}
\usepackage{algpseudocodex}
\usepackage[misc]{ifsym}
\usepackage[pagewise]{lineno}

\theoremstyle{definition}

\usepackage[font={small},labelfont={small}]{caption}

\title{\Large\bf Conformal prediction for frequency-severity modeling}

\author{
{\normalsize Helton Graziadei\footnote{Corresponding author: \Letter \, \tt\footnotesize hltgraziadei@gmail.com}} \\
\textit{\small School of Applied Mathematics, Getulio Vargas Foundation, Rio de Janeiro, Brazil} \bigskip \\ 
{\normalsize Paulo C. Marques F.} \\
\textit{\small Insper Institute of Education and Research, São Paulo, Brazil} \bigskip \\
{\normalsize Eduardo F. L. de Melo} \\ 
\textit{\small School of Applied Mathematics, Getulio Vargas Foundation, Rio de Janeiro, Brazil} \\
\textit{\small SUSEP - Superintendence of Private Insurance, Rio de Janeiro, Brazil} \\
\textit{\small UERJ - State University of Rio de Janeiro, Brazil} \bigskip \\ 
{\normalsize Rodrigo S. Targino} \\
\textit{\small School of Applied Mathematics, Getulio Vargas Foundation, Rio de Janeiro, Brazil} \bigskip 
} 

\date{\footnotesize July 2024}

\begin{document}

\maketitle

\begin{abstract}
\bigskip 
We present a model-agnostic framework for the construction of prediction intervals of insurance claims, with finite sample statistical guarantees, extending the technique of split conformal prediction to the domain of two-stage frequency-severity modeling. The framework effectiveness is showcased with simulated and real datasets using classical parametric models and contemporary machine learning methods. When the underlying severity model is a random forest, we extend the two-stage split conformal prediction algorithm, showing how the out-of-bag mechanism can be leveraged to eliminate the need for a calibration set in the conformal procedure.
\end{abstract}


\thispagestyle{empty}

\section{Introduction}\label{sec:intro}

The statistical modeling of insurance claims is a crucial task in the property and casualty insurance industry. An essential element in this modeling process is the two-stage approach, encompassing a frequency model and a severity model. For a given policy, in the first stage, a frequency model predicts the number of associated claims, while in the second stage a severity model predicts the average financial impact or size of the claims. Together, these two models map relevant predictors, such as the policyholder's age, geographical location, history, and behavior, to the response variables describing the frequency and severity of the claims associated with the policy under consideration. This classical two-stage approach, known as the frequency-severity model, has been a perennial staple in the process of risk categorization, premium calculation, and, in a broader context, risk quantification of business portfolios for specific insurance segments \cite{ohlsson,frees,shi}.

Traditionally, generalized linear models \cite{mcculloch} have been the standard choice for both the frequency and the severity stages. A count regression model, typically based on the Poisson or the negative binomial distributions, is employed for the frequency variable, while a continuous regression model is designed for the severity component, commonly based on the gamma or the log-normal distributions. In recent years, we have witnessed the inception of a paradigm shift, with the insurance industry increasingly gravitating towards the use of modern machine learning techniques \cite{su,roel} for the two modeling stages. However, the plethora of currently available machine learning algorithms share a common limitation, almost always focusing on pointwise predictions, without an appropriate quantification of the confidence in the forecasts made through the assorted modeling techniques.

In light of this, our goal is to explore the use of conformal prediction techniques \cite{vovk2005} to quantify the uncertainty in the process of frequency-severity predictive modeling. Conformal prediction is a model-agnostic and universally applicable framework capable of producing prediction intervals with finite sample statistical guarantees. We aim to adapt the split conformal prediction procedure to the specific characteristics of the two-stage frequency-severity modeling process used in the insurance industry. While doing so, we have the opportunity to assess the impact on the efficiency of the conformalization procedure of different choices for the models in the severity stage, with this efficiency being measured by the average width of the prediction intervals generated for a batch of forecasts.

In Section \ref{sec:freqsev}, we discuss the general process of frequency-severity modeling. Section \ref{sec:tsscp} presents the key ideas behind split conformal prediction \cite{vovk1999,vovk2005,lei}, showing how this technique can be extended to the two-stage frequency-severity scenario. The framework is exemplified in Section \ref{sec:examples} using simulated and real datasets, including a novel dataset of crop insurance claims in Brazil compiled by the authors from public sources. In Section \ref{sec:oobext}, we show how the technique introduced in \cite{johansson2014} can be extended to the two-stage split conformal prediction case when the underlying predictive models are random forests, using the out-of-bag mechanism to eliminate the need for a calibration set in the conformal procedure. The Appendix gives pointers to open access software, coded in \texttt{R} \cite{R}, implementing all the examples in the paper. We state our conclusions in Section \ref{sec:concl}.

\subsection*{Related work}

To the best of the authors' knowledge, this is the first application of conformal prediction techniques to uncertainty quantification in the process of frequency-severity predictive modeling.

\section{Frequency-severity modeling}\label{sec:freqsev}

In the actuarial problems examined in this paper, the sample unit will be a policy belonging to a particular policyholder. For each sample unit, a claim is an event for which the policyholder demands economic compensation. The claim frequency is the total number of claims that occurred during the policy term. The claim severity is the average claim cost, that is, the total claim amount divided by the number of claims \cite{ohlsson}. In some cases, we may extend this basic setting, introducing a specific level of data aggregation. For example, in Section \ref{ssec:crop}, a soybean crop insurance dataset from Brazil is analyzed, for which each row aggregates all contracts at the municipality level, extending the meaning of the term ``policyholder''.

A general nonparametric description of the data generating process goes as follows. Let $(X_1,D_1,Y_1),\dots,(X_n,D_n,Y_n),(X_{n+1},D_{n+1},Y_{n+1})$ be a sample of independent and identically distributed triplets, in which, for the $i$-th sample unit, $X_i\in\mathbb{R}^p$ is a vector of predictors, $D_i\in\mathbb{N}$ denotes the frequency of incurred claims, and $Y_i\in\mathbb{R}$ is the corresponding severity. Table \ref{tab:datastr} outlines this data structure for the motor third party liability dataset analyzed in Section \ref{ssec:mtpl}. Data is assumed to be generated in a two-stage fashion, with $D_i \mid X_i \sim F$, for some discrete distribution $F$ supported on the natural numbers, and $Y_i \mid X_i, D_i \sim G$, in which the severity distribution $G$ is such that $Y_i=0$, whenever $D_i=0$.

Predictive modeling in this context is implemented using two regression models applied in succession: a first model predicts the claim frequency from the available predictors, while a second model predicts the claim severity from the available predictors and the claim frequency predicted by the first model. As mentioned above, Poisson and gamma regressions are standard parametric choices for the frequency and severity stages, respectively. More recently, modern machine learning algorithms such as random forests \cite{breimanRF} have been applied successfully in this scenario, with substantial predictive performance gains \cite{su,roel}. One of our goals in the following sections is to contrast the generalization capabilities of the parametric and machine learning approaches, understanding the impacts in the uncertainty quantification obtained through the conformalization of both choices. 

\begin{table}[t!]
\small
\centering
\caption{Outline of the motor third party liability dataset discussed in Section \ref{ssec:mtpl}, showing ten policies and the values of the observed frequency and severity of the claims, as well as the values of two predictors: the age, in years, and the sex of the policyholder.}
\begin{tabular}{cccccc}
\hline\hline
Claim Count ($D_i$) & Total Claim Amount & Severity ($Y_i$) & Age ($X_{i,1}$) & Sex ($X_{i,2}$) \\ 
\hline\hline
1 & 1,618.00 & 1,618.00 & 50 & \texttt{male} \\
2 & 232.60 & 116.30 & 33 & \texttt{male} \\
0 & 0 & 0 & 64 & \texttt{female} \\
1 & 155.97 & 155.97 & 28 & \texttt{female} \\
0 & 0 & 0 & 60 & \texttt{male} \\
0 & 0 & 0 & 26 & \texttt{male} \\
1 & 62.42 & 62.42 & 41 & \texttt{female} \\
0 & 0 & 0 & 58 & \texttt{female} \\
4 & 1,776.03 & 444.01 & 24 & \texttt{male} \\
3 & 1,980.69 & 653.56 & 38 & \texttt{male} \\
\hline\hline 
\end{tabular}
\label{tab:datastr}
\end{table} 

\section{Two-stage split conformal prediction}\label{sec:tsscp}

Conformal prediction \cite{vovk1999,papadopoulos,vovk2005,shafer2008,lei,fontana,angelopoulos,coverage} is a nonparametric method developed in the late 1990s to quantify the confidence in the forecasts made by general predictive models through the construction of prediction intervals with finite sample coverage guarantees. Here, the word ``coverage'' refers to whether a prediction set contains the future observed value of the variable forecasted by a given model or algorithm; not to be confused with the common actuarial term ``type of coverage'' provided by an insurance contract, meaning to which extent of risk or liability the contract protects an individual or entity.

\subsection{The classical procedure}\label{ssec:single}

\noindent Before we show how the conformal prediction techniques can be applied to the two-stage frequency-severity modeling involved in our actuarial setting, we need a brief description of the standard ``single-stage'' split conformal prediction procedure \cite{vovk2005,papadopoulos,lei}. In a regression context, suppose that we have a random sample of independent and identically distributed pairs $(X_1,Y_1),\dots,(X_n,Y_n),(X_{n+1},Y_{n+1})$, in which $X_i\in\mathbb{R}^d$ is a vector of predictors and $Y_i\in\mathbb{R}$ is a response variable. This distributional symmetry assumption can be weakened by assuming only exchangeability among the pairs in the sample. Our goal is to construct a prediction interval for $Y_{n+1}$ based on the first $n$ sample pairs and the vector of predictors $X_{n+1}$. In the split conformal prediction scheme, we randomly split the sample units indexes into two disjoint sets $I_1$ and $I_2$, with $I_1\cup I_2=\{1,\dots,n\}$, corresponding to training and calibration samples, with sizes $n_1$ and $n_2$, respectively. Any regression model $\hat{\mu}$ is constructed using only the pairs in the training sample, and this model is used to compute the calibration sample conformity scores $R_i=|Y_i-\hat{\mu}(X_i)|$, for $i\in I_2$.

One of the key principles in predictive modeling is that we should not rely directly on the data used to build a model -- the training sample -- to gauge the model's generalization capacity. In fact, it is common for contemporary machine learning algorithms to show remarkable in-sample performance and to be able to nearly interpolate the training sample, so that some form of internal regularization mechanism has to be introduced to produce a model with good out-of-sample predictive performance on future data. The first basic intuition behind the split conformal prediction procedure follows this general principle. Since the regression model $\hat{\mu}$ is built without access to any information in the calibration sample, the calibration conformity scores $R_i$ can work as adequate proxies to assess the generalization capacity of the model. The second idea propelling the split conformal prediction procedure is that, since we are modeling our data as independent and identically distributed, we are able to transfer this prediction performance assessment from the calibration sample to unseen data, creating a prediction interval with statistical guarantees for a future response variable $Y_{n+1}$. Formally, this is done as follows.

Let $(R_{(1)}, R_{(2)},\dots,R_{(n_2)})$ denote the ordered calibration conformity scores. Assuming that there are no ties among the conformity scores, a key fact is that, due to the postulated distributional symmetry of the data, the conformity score $R_{n+1}=|Y_{n+1}-\hat{\mu}(X_{n+1})|$ for a future random pair $(X_{n+1},Y_{n+1})$ is ranked uniformly among the ordered calibration conformity scores, meaning that
\begin{equation}\label{eq:uniformity}
  P(R_{n+1}\leq R_{(k)})=k/(n_2+1),
\end{equation}
for $k=1,\dots,n_2$. More precisely, property $(\ref{eq:uniformity})$ follows from the fact that the distribution of the random vector of conformity scores $(R_{i_1},R_{i_2}\dots,R_{i_{n_2}},R_{n+1})$, for $i_1,i_2,\dots,i_{n_2}\in I_2$, is exchangeable. For a real number $t$, let $\lceil t\rceil=\min\{z\in\mathbb{Z}:t\leq z\}$ denote the ceiling of $t$, that is, the smallest integer greater than or equal to $t$. Fix some nominal miscoverage level $0<\alpha<1$, whose value will control the probability that the prediction interval contains the value of the future response variable $Y_{n+1}$, and suppose that the chosen nominal miscoverage level is such that $\lceil(1-\alpha)(n_2+1)\rceil\leq n_2$. Given the arithmetical fact that $(1-\alpha)(n_2+1)\leq\lceil(1-\alpha)(n_2+1)\rceil<(1-\alpha)(n_2+1)+1$, if we choose $k=\lceil(1-\alpha)(n_2+1)\rceil$ in equation (\ref{eq:uniformity}), we obtain the coverage property
\begin{equation}\label{eq:mvp}
  1 - \alpha \leq P(Y_{n+1} \in C^{(1-\alpha)}(X_{n+1})) < 1 - \alpha + \frac{1}{n_2+1},
\end{equation}
in which we defined the conformal prediction interval
\begin{equation}\label{eq:scpint}
  C^{(1-\alpha)}(X_{n+1}) = [\,\hat{\mu}(X_{n+1}) - \hat{r}, \hat{\mu}(X_{n+1}) + \hat{r}\,],
\end{equation}
with $\hat{r}=R_{(\lceil(1-\alpha)(n_2+1)\rceil)}$.

\begin{figure}[t!]
\centering
\includegraphics[width=16cm]{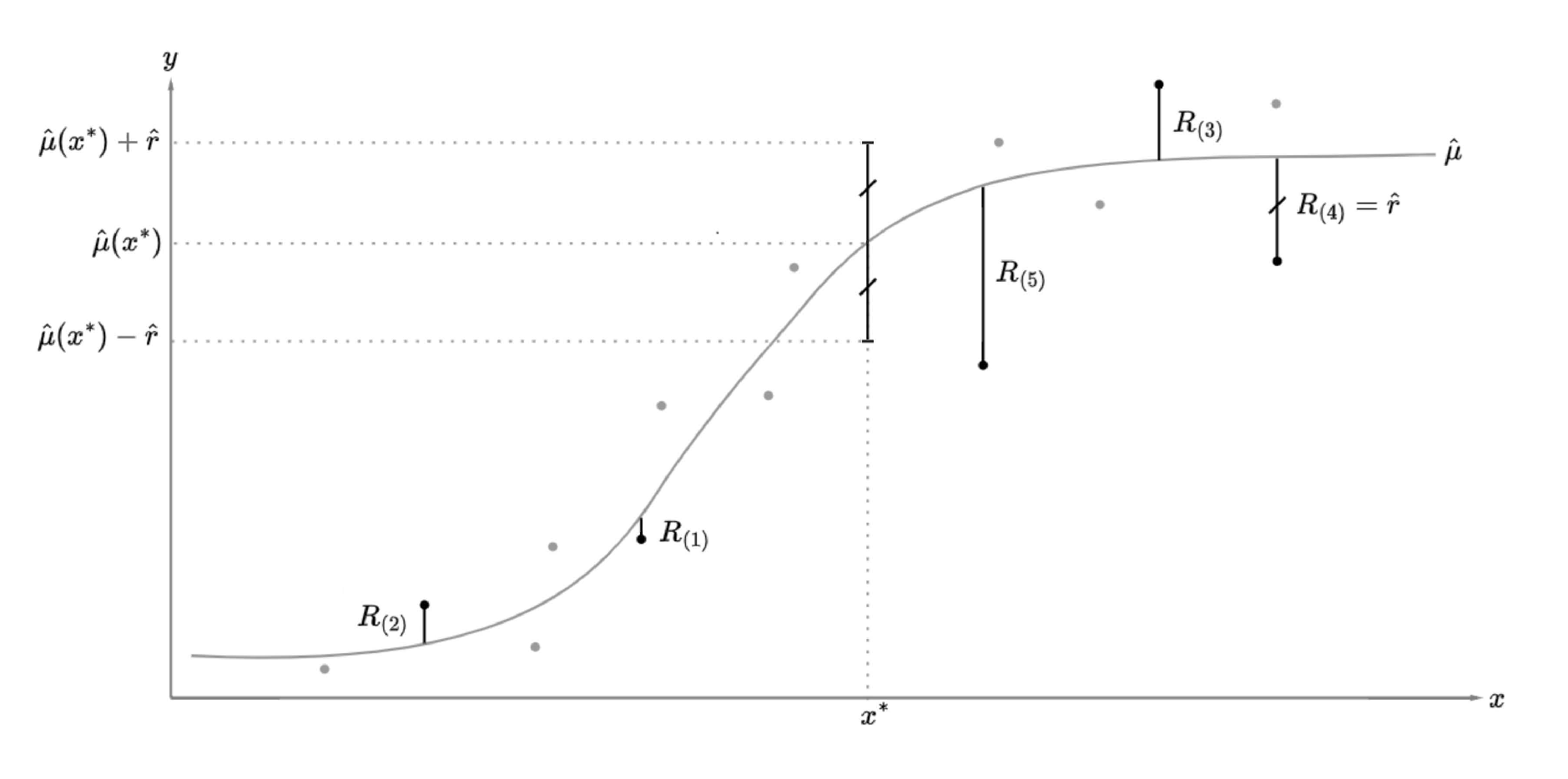}
\caption{An illustration of the split conformal procedure for regression with a single predictor. We start with fourteen data points, which are randomly split into the nine gray and five black points in the figure, representing the training and calibration samples, respectively. The gray line $\hat{\mu}$ is a predictive model fit to the training sample using a nonparametric method. The lengths of the black segments are the values of the calibration conformity scores. The nominal miscoverage level $\alpha=40\%$, so that $\hat{r}=R_{(\lceil(1-0.4)((5+1)\rceil)}=R_{(4)}$. A future predictor $x^*$ and the corresponding conformal prediction interval are depicted in the figure.}
\label{fig:scp}
\end{figure}

It is worth emphasizing that the coverage property $(\ref{eq:mvp})$ is essentially a consequence of the distributional symmetry assumption of independent and identically distributed data and the fact that the predictive model $\hat{\mu}$ is trained without access to the calibration sample information. Furthermore, property $(\ref{eq:mvp})$ holds for every calibration sample size $n_2$, and for data with high dimensional predictors ($d\gg 1$), no matter what model or algorithm is used to construct $\hat{\mu}$. All these combined features account for the so-called universality of the split conformal prediction framework. This split conformal prediction procedure is also known in the literature as the inductive case of conformal prediction \cite{vovk2005}. Figure \ref{fig:scp} provides a visual intuition for the split conformal prediction procedure in a simple regression case with a single predictor variable.

Conformal prediction intervals built from expression $(\ref{eq:scpint})$ for a batch of $m\geq 1$ future vectors of predictors $X_{n+1},\dots,X_{n+m}$ have the same fixed width ($2\times \hat{r}$). To increase the flexibility of the procedure, making it more adaptable to future data, we can use the locally-weighted conformity score introduced in \cite{papadopoulos2010} and discussed in \cite{lei}. The idea is to compute the training sample absolute residuals $\Delta_i=|Y_i-\hat{\mu}(X_i)|$, for $i\in I_1$, and to use $\{(X_i,\Delta_i)\}_{i\in I_1}$ to construct a regression model $\hat{\sigma}:\mathbb{R}^d\to\mathbb{R}_+$. The intuition is that this second regression model measures the variability of the predictions made by $\hat{\mu}$, allowing us to define the new locally-weighted calibration conformity scores $R_i=|Y_i-\hat{\mu}(X_i)|/\hat{\sigma}(X_i)$, for $i\in I_2$. If we repeat the split conformal prediction argument presented above using these locally-weighted conformity scores, we are led to a new expression for the conformal prediction interval
$$
  C^{(1-\alpha)}(X_{n+1}) = [\, \hat{\mu}(X_{n+1}) - \hat{r}\times\hat{\sigma}(X_{n+1}), \hat{\mu}(X_{n+1}) + \hat{r}\times\hat{\sigma}(X_{n+1}) \,],
$$
again satisfying the coverage property (\ref{eq:mvp}), but now the conformal prediction intervals have adaptive width.

\subsection{Two-stage setting}\label{ssec:twostage}

\noindent To adapt the split conformal prediction to the two-stage setting, we have to modify the classical split conformal prediction procedure in order to account for the presence of the frequency and the severity models, as well as a third severity variability model, which will be necessary to construct the denominator of the locally-weighted conformity score appropriate to this new context. In this extension, we aim to preserve the essential exchangeability of the conformity scores, from which the coverage property $(\ref{eq:mvp})$ follows.

We start with the random sample $(X_1,D_1,Y_1),\dots,(X_n,D_n,Y_n),(X_{n+1},D_{n+1},Y_{n+1})$ described in Section \ref{sec:freqsev}. This random sample's first $n$ triplets are split into training and calibration samples as described in the classical split conformal prediction procedure. Again, we denote the index set for the training sample units by $I_1$, and the index set for the calibration sample units by $I_2$, and use $n_2$ to denote the calibration sample size. A frequency model $\hat{\mu}$ is built using the information in the training sample units $(X_i,D_i)$, for $i\in I_1$. After this, a severity model $\hat{\psi}$ is built from $(X_i,D_i,Y_i)$, using only those training sample units $i\in I_1$ for which the observed frequency $D_i>0$. Using the severity model $\hat{\psi}$, we compute the training sample absolute residuals $\Delta_i=|Y_i-\hat{\psi}(X_i)|$, for $i\in I_1$, and a severity variability model $\hat{\sigma}:\mathbb{R}^d\to\mathbb{R}$ is built from from $(X_i,\Delta_i)$, considering only the training sample units $i\in I_1$ such that the observed frequency $D_i>0$.

Once we have trained the frequency $\hat{\mu}$, severity $\hat{\psi}$, and severity variability $\hat{\sigma}$ regression models, we have the means to define and compute the calibration sample conformity scores $R_i=|Y_i-\hat{\psi}(X_i,\hat{\mu}(X_i))|/\hat{\sigma}(X_i,\hat{\mu}(X_i))$, for $i\in I_2$. As before, it follows from the exchangeability of the conformity scores that the coverage property $(\ref{eq:mvp})$ holds if we define the two-stage conformal prediction interval for a future severity $Y_{n+1}$ by
$$
  C^{(1-\alpha)}(X_{n+1}) = [\,\max\,\{0,\hat{\psi}(X_{n+1},\hat{\mu}(X_{n+1})) - \epsilon\}, \hat{\psi}(X_{n+1},\hat{\mu}(X_{n+1})) + \epsilon\,],
$$
in which $X_{n+1}$ is a future vector of predictors and $\epsilon=R_{(\lceil(1-\alpha)(n_2+1)\rceil)} \times \hat{\sigma}(X_{n+1},\hat{\mu}(X_{n+1}))$. A formal description of this two-stage split conformal prediction procedure is given in Algorithm \ref{algo:tsscp}. For a specified nominal miscoverage level $\alpha$, the coverage property \ref{eq:mvp} of the conformal prediction interval produced by Algorithm \ref{algo:tsscp} follows from the distributional symmetry of the triples $(X_i,D_i,Y_i)$, by the same reasoning presented in Section \ref{ssec:single}.

In the next section, we run experiments with synthetic and real datasets varying the choices of the severity and severity variability models between a traditional parametric choice and a contemporary machine learning algorithm. Of course, the coverage property $(\ref{eq:mvp})$ is model-agnostic and holds for both cases. However, this arrangement allows us to evaluate the impact of each alternative on the predictive performance, as measured by the averages of the prediction intervals widths produced by the two-stage split conformal prediction procedure in both cases, for different datasets.

\begin{algorithm}[t!]
\caption{Two-stage split conformal prediction}\label{algo:tsscp}
\begin{algorithmic}[1]
  \Require Dataset $\{(x_i,d_i,y_i)\}_{i=1}^n$, training and calibration indexes $I_1$ and $I_2$, calibration sample size $n_2=|I_2|$, future vector of predictors $x_{n+1}\in\mathbb{R}^p$, and nominal miscoverage level $0<\alpha<1$.
  \Ensure Prediction interval.
  \Statex
  \State Train frequency model $\hat{\mu}$ from $\{(x_i,d_i):i\in I_1\}$
  \State $I_1^+ \gets \{i\in I_1: d_i>0\}$
  \State Train severity model $\hat{\psi}$ from $\{(x_i,d_i,y_i):i\in I_1^+\}$
  \For{$i\in I_1^+$}
    \State $\delta_i \gets |y_i - \hat{\psi}(x_i,d_i)|$
  \EndFor
  \State Train severity variability model $\hat{\sigma}$ from $\{(x_i,d_i,\delta_i):i\in I_1^+\}$
  \For{$i\in I_2$}
    \State $r_i \gets |y_i - \hat{\psi}(x_i,\hat{\mu}(x_i))| / \hat{\sigma}(x_i,\hat{\mu}(x_i))$
  \EndFor
  \State $\epsilon \gets r_{(\lceil(1-\alpha)(n_2+1)\rceil)} \times \hat{\sigma}(x_{n+1},\hat{\mu}(x_{n+1}))$
  \State \Return{$[\,\max\,\{0,\hat{\psi}(x_{n+1},\hat{\mu}(x_{n+1})) - \epsilon\}, \hat{\psi}(x_{n+1},\hat{\mu}(x_{n+1})) + \epsilon\,]$}
\end{algorithmic}
\end{algorithm}

\section{Synthetic and real datasets}\label{sec:examples}

In this section, we apply the two-stage split conformal prediction procedure to three different datasets, varying the kind of models used for the second stage in the frequency-severity configuration. A more traditional choice of gamma regressions for the severity stage works as a benchmark against an alternative scheme using random forest \cite{breimanRF}. We begin with a simulated dataset and move to a motor third party liability dataset from the Belgian market. The last analysis involves a new dataset of Brazilian crop insurance. Throughout the analysis, we register the observed coverages, in agreement with property $(\ref{eq:mvp})$, and discriminate the performance of the models regarding the averages of the prediction intervals widths produced by the two-stage split conformal prediction procedure on each setup.

\subsection{Synthetic data}\label{ssec:synth}

\begin{figure}[t!]
\centering
\includegraphics[width=16cm]{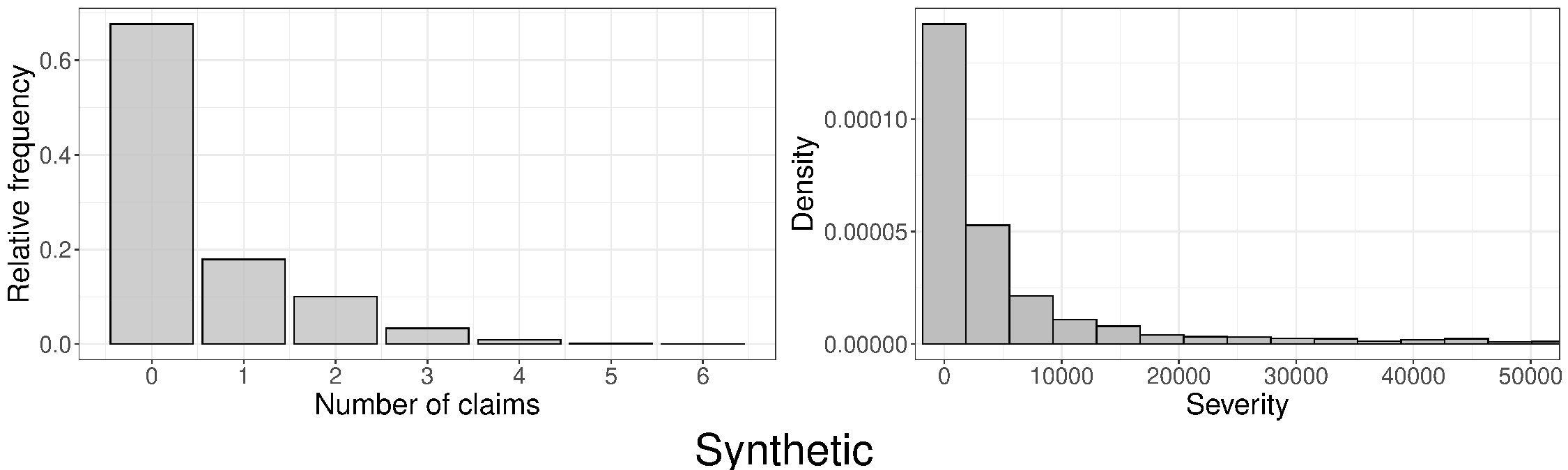}
\caption{Frequency and severity distributions for the 10,000 sample units in the synthetic dataset.}
\label{fig:syntheda}
\end{figure}

Following the structure of the two-stage data generating process described in Section \ref{sec:freqsev}, we simulate a size 10,000 dataset with ten independent predictors $X_{i,1},X_{i,2},\dots X_{i,10}$, each one with $U[0,10]$ distribution. The frequency variable $D_{i}$ is drawn from a mixture, with symmetric weights, comprising a point mass at zero and a Poisson($\lambda_i$) distribution, with rate parameter $\lambda_i = e^{0.01\times X_{1,i}}$. The purpose of this mixture is to emulate the behavior of the zero-inflated frequencies commonly found in real insurance datasets. The severity variable $Y_i$ is sampled conditionally: if $D_i=0$, then $Y_i=0$, otherwise, we draw $Y_i$ from an exponential distribution with expectation $4\times e^{X_{i,2}} + \sin(X_{i,3}\times X_{i, 4}) + 5\times X_{i, 5}^3$. Figure \ref{fig:syntheda} depicts the distributions of the claim frequency ($D_i$) and severity ($Y_i$). Claims with zero frequency ($D_i = 0$) comprise a significant portion (67.43\%) of the data. Notice that in this data generating process, the last five predictors are unrelated to $D_i$ and $Y_i$. These predictors are introduced only to create a degree of sparsity in the simulated dataset.

\begin{figure}[t!]
\centering
\includegraphics[width=15cm]{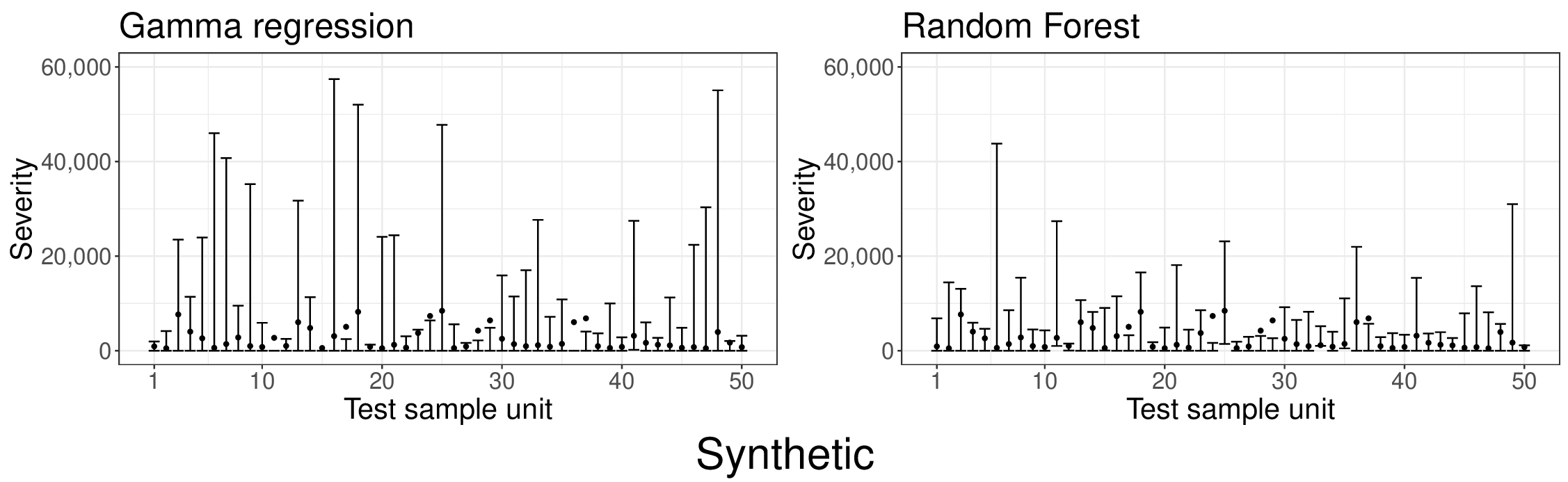}
\caption{Prediction intervals produced by Algorithm \ref{algo:tsscp} for fifty test sample units in the synthetic dataset, using a nominal miscoverage level $\alpha=10\%$. The black dots are the observed severity values. On the left and right figures, we have the results using gamma regressions and random forests, respectively, for the severity stage.}
\label{fig:synthsplit}
\end{figure}

\begin{figure}[t!]
\centering
\includegraphics[width=14cm]{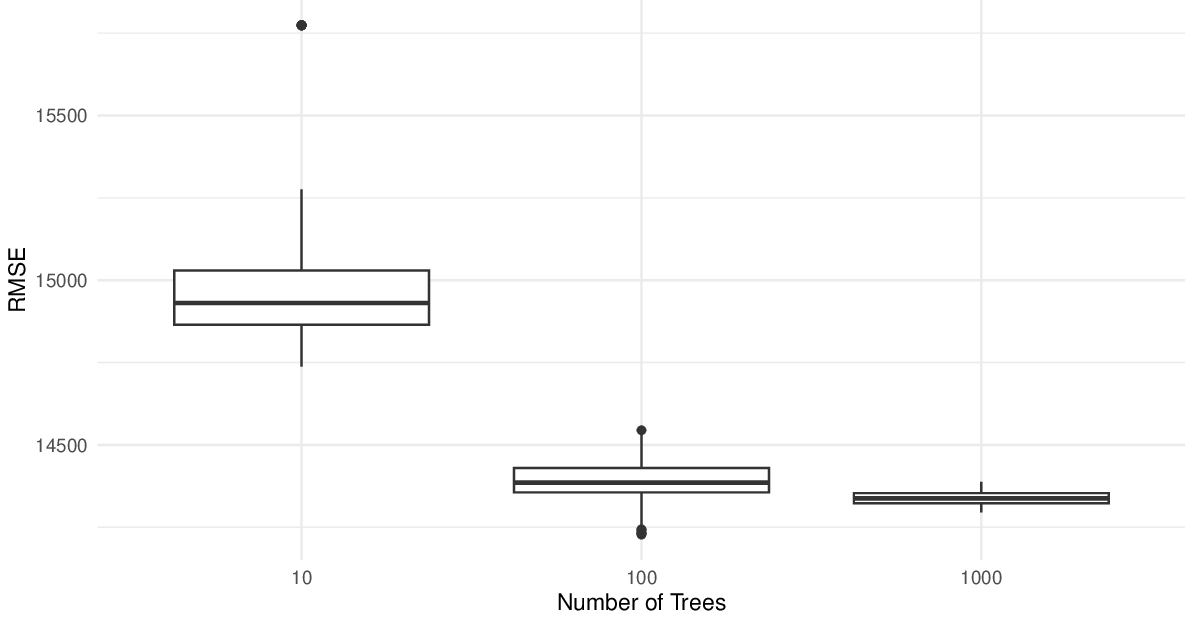}
\caption{RMSE distributions based on 100 replications of the random forest models with 10, 100, and 1,000 trees.}
\label{fig:synth_rmse}
\end{figure}

This synthetic dataset is randomly partitioned into a training sample of size 5,000, a calibration sample of size 2,500, and a test sample of size 2,500. We use a random forest with 1,000 trees as the predictive model for the claim frequency. For the severity and severity variability models, we consider two alternatives: gamma regressions and random forests with 1,000 trees. For the random forest, we assess the impact of the number of trees on the predictive performance to motivate the use of an ensemble with 1,000 trees. Figure \ref{fig:synth_rmse} shows the root mean squared error (RMSE) distributions obtained from 100 replications of the random forest model trained with 10, 100, and 1,000 trees. The results show that while increasing the number of trees from 10 to 100 substantially reduces the median RMSE and its variability, the improvement from 100 to 1,000 trees is modest but still noticeable, stabilizing the predictions. We also report the running times associated with each model configuration. Table \ref{tab:running_times} reports descriptive statistics of the training times (in milliseconds) to fit the random forest models with 10, 100, and 1,000 trees, using 100 replications. Even for the random forest model with 1,000 trees, the median computation time remains around 10 milliseconds, indicating that predictive stability can be achieved with manageable computational cost. Regarding test set predictive performance, the root mean squared error for the predicted severities using the gamma regressions and the random forests schemes are 23,775.38 and 14,397.69, respectively. 

\begin{table}[t!]
\caption{Running times (in milliseconds) to train the random forest models with 10, 100, and 1,000 trees, for the synthetic dataset. Reported values correspond to median, minimum, and maximum running times, using 100 replications.}
\centering
\begin{tabular}{ccccc}
\hline\hline
Number of Trees & Median (ms) & Min (ms) & Max (ms) \\
\hline\hline
10              & 0.10      & 0.09   & 0.47  \\
100             & 0.68      & 0.40   & 0.88   \\
1,000            & 7.72      & 3.97   & 10.80 \\ 
\hline\hline
\end{tabular}
\label{tab:running_times}
\end{table}

Using a nominal miscoverage level $\alpha=10\%$, Figure \ref{fig:synthsplit} shows the prediction intervals for fifty test sample units in the synthetic dataset. Using gamma regressions for the severity and severity variability models, the average prediction interval width is 23,717.68, while the random forests alternative gives prediction intervals with lower average width equal to 12,161.26. The observed coverages of the prediction intervals considering the full test sample are 90.25\% and 90.29\%, for the gamma regressions and the random forests cases, respectively.

\subsection{Motor third party liability in Belgium}\label{ssec:mtpl}

\begin{figure}[t!]
\centering
\includegraphics[width=16cm]{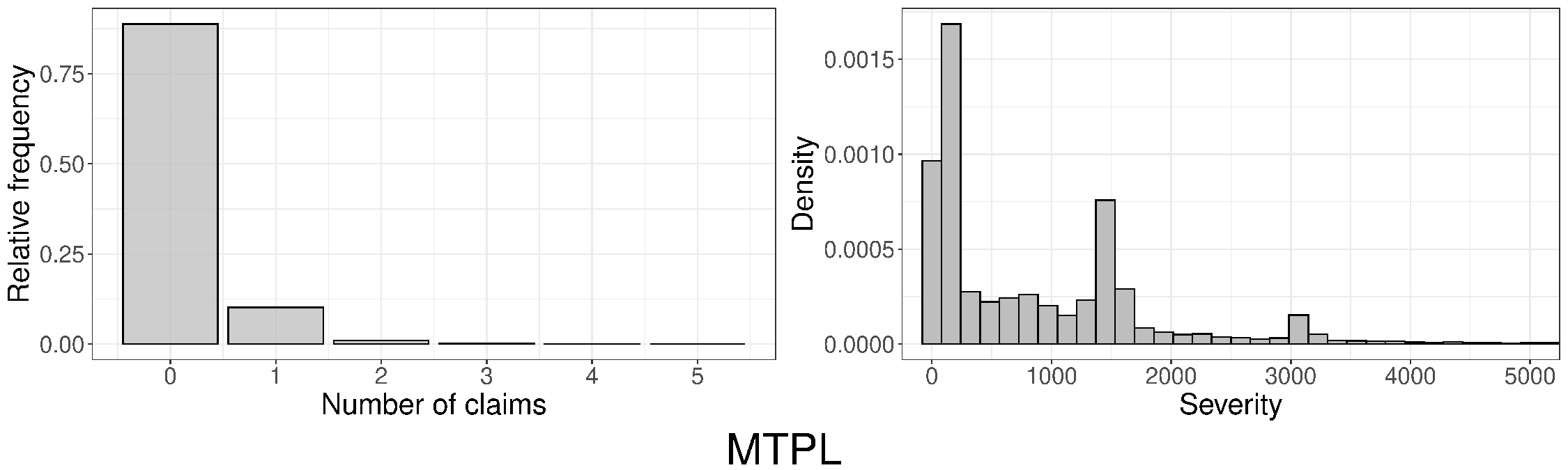}
\caption{Frequency and severity distributions for the 163,212 policies in the motor third party liability dataset.}
\label{fig:mtpleda}
\end{figure}

In this example, we use a motor third party liability (MTPL) portfolio dataset from a Belgian insurance company, related to the year 1997, also analyzed in \cite{roel,denuit,klein}. This dataset contains information about 163,212 policyholders. A description of the variables available in the dataset is given in Table \ref{tab:mtpl}. The proportion of zero claims contracts in this dataset is approximately 88.80\%. Figure \ref{fig:mtpleda} depicts the distributions of claim frequency and severity for the MTPL dataset. 

\begin{table}[t!]
\small
\centering
\captionof{table}{Variables in the motor third party liability dataset (reproduced from \cite{roel}).}\label{tab:mtpl}
\begin{tabular}{rl}
\hline\hline
Variable & Description \\
\hline\hline
\texttt{Type} & Type of insurance plan \\
\texttt{Fuel} & Motor fuel of the vehicle \\ 
\texttt{Sex} & Sex of the policyholder (male or female) \\
\texttt{Use} & Main use of the vehicle (private or work) \\ 
\texttt{Fleet} & Indicates whether the vehicle is part of a fleet \\ 
\texttt{Ageph} & Age of the policyholder (years) \\
\texttt{Power}  & Horsepower of the motor vehicle \\
\texttt{Bm} & \textit{Bonus malus} scale (lower value indicates better claim history)  \\ 
\texttt{Lat} & Latitude of the center of the municipality where the policyholder resides \\ 
\texttt{Long} & Longitude of the center of the municipality where the policyholder resides \\ 
\texttt{NClaims} & The number of claims \\ 
\texttt{Amount} & The total amount claimed by the policyholder (euros) \\ 
\texttt{Expo} & Contract period (fraction of a year) \\
\hline\hline
\end{tabular}
\end{table}

The MTPL dataset is randomly partitioned into training, calibration, and test samples, according to the proportions 50\%, 25\%, and 25\%, respectively. Again, we use a random forest with 1,000 trees as the predictive model for the claim frequency. As before, we consider two alternatives for the severity and severity variability models: gamma regressions and random forests with 1,000 trees. For the severity predictions on the test set, the root mean square error using the gamma regressions and the random forests are 1,709.01 and 1,210.45 euros, respectively.

\begin{figure}[t!]
\centering
\includegraphics[width=15cm]{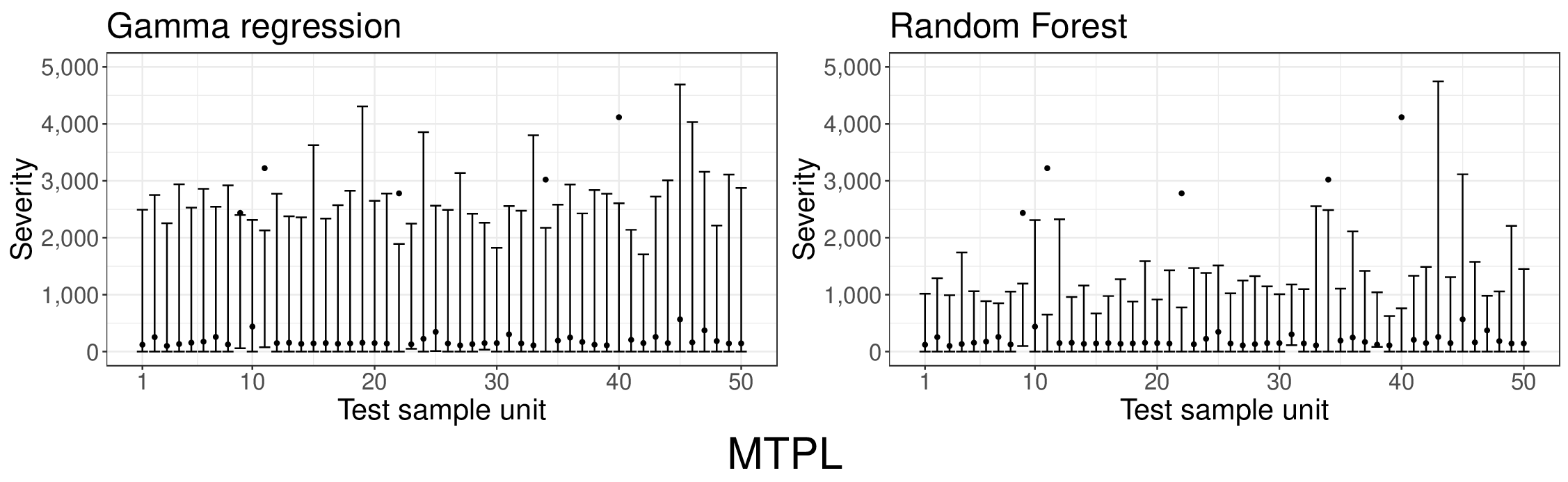}
\caption{Prediction intervals produced by Algorithm \ref{algo:tsscp} for fifty test sample units in the motor third party liability dataset, using a nominal miscoverage level $\alpha=10\%$. The black dots are the observed severity values. On the left and right figures, we have the results using gamma regressions and random forests, respectively, for the severity stage.}
\label{fig:mtplsplit}
\end{figure}

Using a nominal miscoverage level $\alpha=10\%$, Figure \ref{fig:mtplsplit} shows the prediction intervals for fifty test sample units. For the scenario using the gamma regressions models, the average prediction interval width is 2,902.11 euros, while the random forests scheme gives prediction intervals with lower average width equal to 1,334.31 euros. The observed coverages of the prediction intervals are 89.70\% and 89.93\%, for the gamma regressions and the random forests cases, respectively.

\subsection{Crop insurance in Brazil}\label{ssec:crop}

In this last example, we consider data related to crop insurance in Brazil. The data was compiled from the Brazilian Agricultural Ministry public records (raw data is publicly available at \texttt{https://dados.agricultura.gov.br/dataset/sisser3}), encompassing crop insurance policies within a time frame ranging from March 2016 to February 2022. All policies belong to the Rural Grant Program ({\it Programa de Subvenção Rural}), maintained by the Brazilian federal government to subsidize a percentage of the insurance premiums for Brazilian farmers.

In our analysis, we focus on policies related to soybean crops due to the peculiarities of their agricultural cycles and relevance to the Brazilian economy (around 42\% of the whole Brazilian agriculture production). The dataset is aggregated at the municipality level. We consider the sample unit to be the municipality in a given harvest year, giving us a total of 9,507 sample units in this dataset. Furthermore, municipalities without at least one policy during a specific agricultural year were excluded from the dataset. The dataset incorporates relevant climate variables such as monthly cumulative precipitation (in millimeters) and monthly average temperatures (in Celsius), which were obtained from the Climatic Research Unit gridded Time Series \cite{harris2020}, version 4. We use Principal Components Analysis \cite{esl} to reduce the dimensionality of the temperature and the precipitation data. The list of variables available for this crop insurance dataset is presented in Table \ref{tab:crop}.

\begin{table}[t!]
\small
\centering
\captionof{table}{Variables in the crop insurance dataset.}\label{tab:crop}
\begin{tabular}{rl}
\hline\hline
Variable & Description \\
\hline\hline
\texttt{Municipality} & Name of the municipality \\
\texttt{Year} & Agricultural year (1 to 6) \\
\texttt{Latitude} & Latitude of the municipality \\
\texttt{Longitude} & Longitude of the municipality \\ 
\texttt{AWC} & Available water capacity of the municipality \\
\texttt{Soil} & Predominant type of soil found in the municipality \\ 
\texttt{Area} & The acreage of insured crops (hectares) \\ 
\texttt{Irrigation} & The proportion of irrigated agricultural land \\ 
\texttt{TempPC1} & Previous year temperature score on the first principal component \\ 
\texttt{TempPC2} & Previous year temperature score on the second principal component \\ 
\texttt{PrecPC1} & Previous year precipitation score on the first principal component \\ 
\texttt{PrecPC2} & Previous year precipitation score on the second principal component \\ 
\texttt{PrecPC3} & Previous year precipitation score on the third principal component \\ 
\texttt{PrecPC4} & Previous year precipitation score on the fourth principal component \\ 
\texttt{Claims} & Number of claims \\
\texttt{RelativeLoss} & Severity (Brazilian reais per claim) \\ 
\hline\hline
\end{tabular}
\end{table}

\begin{figure}[t!]
\centering
\includegraphics[width=16cm]{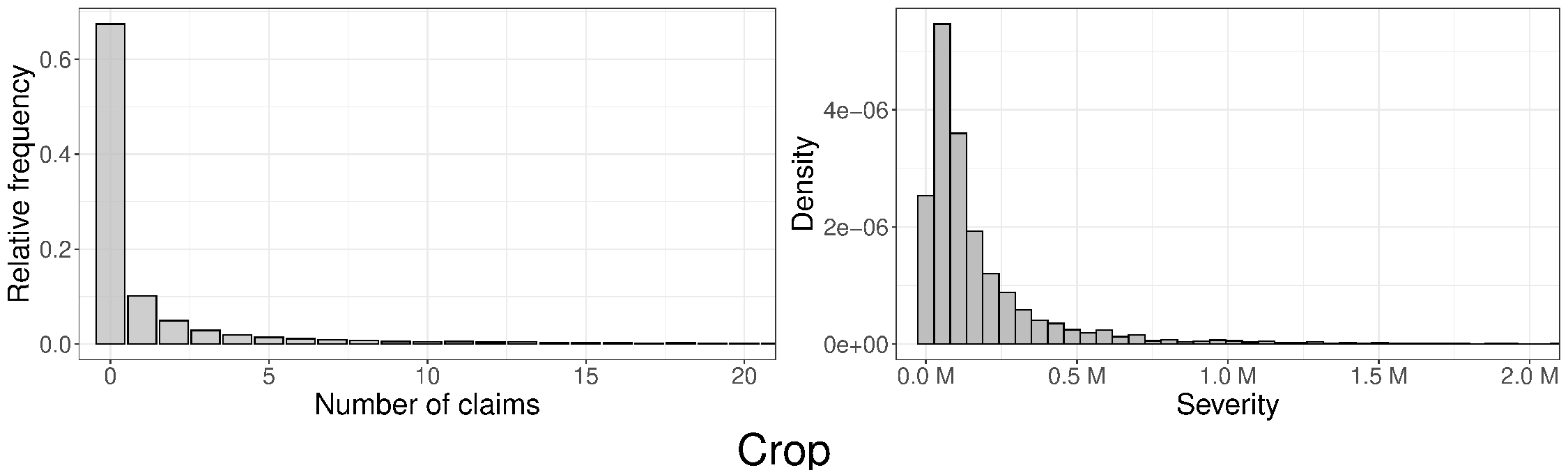}
\caption{Frequency and severity distributions for the 9,507 sample units in the crop insurance dataset.}
\label{fig:cropeda}
\end{figure}

The claim frequency and severity distributions are given in Figure \ref{fig:cropeda}. The proportion of sample units with nonzero claims in this dataset is approximately 32.57\%. This substantially high proportion is due to soybean crops being highly susceptible to climate factors, especially droughts, which occurred in the considered time frame. 

We randomly split the crop insurance dataset into training, calibration, and test samples, with proportions 50\%, 25\%, and 25\%, respectively. As before, we use a random forest with 1,000 trees as the predictive model for the claim frequency. Following the arrangements in the previous examples, we consider two alternatives for the severity and severity variability models: gamma regressions and random forests with 1,000 trees. For the severity predictions on the test set, the root mean square error using the gamma regressions and the random forests setups are 281,478.10 and 177,601.50, respectively.

\begin{figure}[t!]
\centering
\includegraphics[width=16cm]{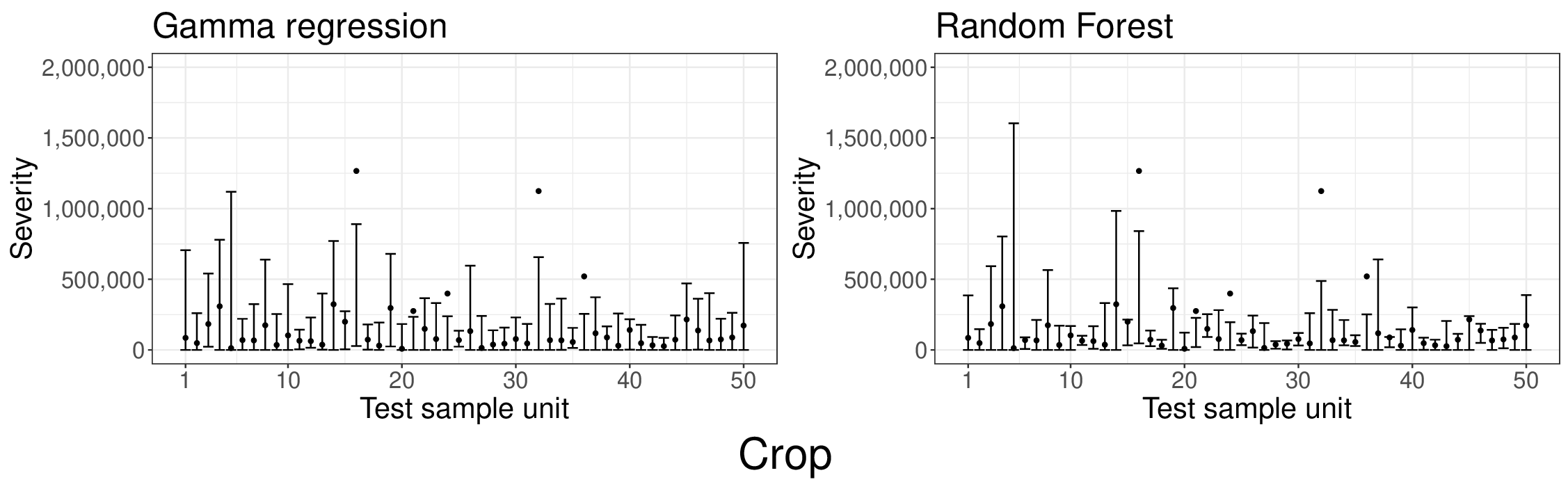}
\caption{Prediction intervals produced by Algorithm \ref{algo:tsscp} for fifty test sample units in the crop insurance dataset, using a nominal miscoverage level $\alpha=10\%$. The black dots are the observed severity values. On the left and right figures, we have the results using gamma regressions and random forests, respectively, for the severity stage.}
\label{fig:cropsplit}
\end{figure}

Using a nominal miscoverage level $\alpha=10\%$, Figure \ref{fig:cropsplit} shows the prediction intervals for fifty test sample units. For the scenario using the gamma regressions, the average prediction interval width is 459,556.80, while the random forests scheme gives a lower average prediction interval width of 349,026.40. The observed coverages of the prediction intervals are 90.71\% and 90.50\%, for the gamma regressions and the random forests setups, respectively.

\section{Out-of-bag extension}\label{sec:oobext}

Our discussion in Section \ref{sec:tsscp} emphasized that in the split conformal prediction procedure the calibration sample cannot be used in the training process of the underlying predictive model, which is built from the training sample information. In a sense, the necessary presence of the calibration sample in the split conformal prediction procedure becomes a nuisance in terms of predictive performance: if by any means we were able to eliminate the necessity for this data splitting process, building our predictive model with all the available data, we would end up with a model with better generalization capacity. When random forests are used for the severity stage, as we did in the examples of Section \ref{sec:examples}, the bootstrap process used in the random forest training opens the possibility of exploiting the so-called out-of-bag mechanism to emulate the existence of a calibration sample, enabling us to use the entire available information to create our predictive model. For the single-stage case, this possibility was investigated for the first time in the work of Johansson et al. \cite{johansson2014} (see also \cite{bostrom2022}). In Section \ref{ssec:rfoob}, we review the details of the training process of a random forest and establish the necessary notations for our development. In Section \ref{ssec:tsoobcp}, we give an out-of-bag extension of the two-stage split conformal procedure, comparing its efficiency when applied to the three datasets already analyzed in Section \ref{sec:examples} with the data splitting methods.

\subsection{Random forests and the out-of-bag mechanism}\label{ssec:rfoob}

In the usual supervised learning setting \cite{esl}, Breiman and coauthors \cite{cart} developed in the 1980s the classification and regression trees (CART) algorithm, which recursively partitions the space of predictor variables, greedily looking for splits in the training data that minimize, in the regression case, a quadratic cost function. Resampling a size $n$ training set by drawing $n$ observations uniformly with replacement, we get a bootstrap sample \cite{efron} that can be used to train a tall regression tree using the CART algorithm. Repeating this process $B$ times, we have an ensemble of regression trees $\hat{\psi}^{(1)},\dots,\hat{\psi}^{(B)}$, whose predictions can be averaged to produce the aggregated regression function
$$
  \hat{\psi}(\;\cdot\;) = \frac{1}{B} \sum_{j=1}^B\hat{\psi}^{(j)}(\;\cdot\;).
$$
This general aggregation process of regression functions trained from bootstrap samples, known as bagging \cite{breimanBAG}, can be optimized, in a sense to be discussed further ahead, by uniformly drawing without replacement a random subset of predictors when deciding each split for each regression tree in the ensemble. The resulting regression function is called a random forest \cite{breimanRF}.

Since its inception, random forests have consistently demonstrated remarkable generalization performance in numerous supervised learning tasks and have become a permanent staple in predictive modeling. This generalization capacity of random forests can be understood from the classical bias-variance trade-off decomposition \cite{esl}, which states that the expected generalization error of any regression method can be factored into a squared bias and a variance component. In general, a tall individual regression tree trained with enough data has low bias and high variance, in the sense that it can approximate well a rich class of functions (low bias), but small changes in the training sample may modify the resulting regression tree substantially (high variance). By averaging the predictions of a large number of tall regression trees, random forests effectively create a variance reduction mechanism. Furthermore, Breiman's randomized split decision mechanism, mentioned above, improves the predictive performance by breaking up the correlations among predictions made by the individual regression trees in the ensemble -- which were trained by bootstrapping the same observed training sample -- since the existence of such correlations can get in the way of the variance reduction process.

When bootstrapping a training sample of size $n$, the probability that the $i$-th training sample unit is not included in the bootstrap sample is equal, by independence, to $(1~-~1/n)^n$, which, for large $n$, is approximately equal to $e^{-1}\approx36.8\%$. Consequently, when training a random forest with $B$ regression trees, the $i$-th sample unit will not be used in the training process of approximately $36.8\% \times B$ regression trees in the ensemble. We say the $i$-th sample unit stayed ``out-of-bag'' for the aforementioned regression trees. This out-of-bag mechanism effectively creates for each training sample unit a random sub-forest $\{\hat{\psi}^{(j)}:j\in\mathcal{O}_i\}$, in which $\mathcal{O}_i\subset\{1,2,\dots,B\}$ denote the indexes of the regression trees for which the $i$-th training observation stayed out-of-bag of the corresponding bootstrap samples. Table \ref{tab:oob} illustrates the general idea behind this out-of-bag mechanism.

Although at first sight the out-of-bag mechanism seems to be just a simple side effect of the random forest training process, it can be used to give almost computationally free estimates of the expected generalization error of the model \cite{breimanBAG}, as well as confidence intervals for this quantity \cite{marques}. The general idea is that predictions made for each training sample unit averaging only the regression trees on the corresponding out-of-bag random sub-forest can be used to approximate out-of-sample quantities, such as test errors, and, in the case of conformal prediction, to emulate the behavior of conformity scores. In the next section, we discuss how we can use the out-of-bag mechanism to create an extension of Algorithm \ref{algo:tsscp} that does not involve a calibration sample. The main benefit is that our predictive models can now be constructed using a larger training sample, which will positively impact on their predictive performances.

{\renewcommand{\arraystretch}{1.4}
\begin{table}[t!]
\footnotesize
\centering
\caption{An illustration of the out-of-bag mechanism in action. We have ten training sample units and twenty-five bootstrap samples. The first column gives the indices of the training sample units included on each bootstrap sample obtained by uniformly sampling with replacement the training sample. In the second column, we have the indices of the training sample units which stayed out-of-bag for each bootstrap sample on the first column. The third column represents the regression trees trained from the bootstrap samples on the first column. For the fourth sample unit, the eight boxed regression trees constitute the random sub-forest for which this fourth sample unit stayed out-of-bag of the corresponding bootstrap samples.}
\begin{tabular}{l|l|c}
\hline\hline
{\normalsize Bootstrap sample} & {\normalsize Out-of-bag} & {\normalsize Regression tree} \\
\hline\hline
$7 \hfill\quad\hfill 4 \hfill\quad\hfill 10 \hfill\quad\hfill 8 \hfill\quad\hfill 10 \hfill\quad\hfill 10 \hfill\quad\hfill 4 \hfill\quad\hfill 7 \hfill\quad\hfill 2 \hfill\quad\hfill 3$ & $1 \quad 5 \quad 6 \quad 9$ & $\hat{\psi}^{(1)}$ \\
$5 \hfill\quad\hfill 3 \hfill\quad\hfill 6 \hfill\quad\hfill 6 \hfill\quad\hfill 9 \hfill\quad\hfill 2 \hfill\quad\hfill 1 \hfill\quad\hfill 3 \hfill\quad\hfill 10 \hfill\quad\hfill 4$ & $7 \quad 8$ & $\hat{\psi}^{(2)}$ \\
$1 \hfill\quad\hfill 10 \hfill\quad\hfill 2 \hfill\quad\hfill 2 \hfill\quad\hfill 10 \hfill\quad\hfill 6 \hfill\quad\hfill 5 \hfill\quad\hfill 5 \hfill\quad\hfill 6 \hfill\quad\hfill 2$ & $3 \quad \boxed{4} \quad 7 \quad 8 \quad 9$ & $\boxed{\hat{\psi}^{(3)}}$ \\
$9 \hfill\quad\hfill 1 \hfill\quad\hfill 4 \hfill\quad\hfill 2 \hfill\quad\hfill 8 \hfill\quad\hfill 6 \hfill\quad\hfill 3 \hfill\quad\hfill 1 \hfill\quad\hfill 6 \hfill\quad\hfill 10$ & $5 \quad 7$ & $\hat{\psi}^{(4)}$ \\
$10 \hfill\quad\hfill 10 \hfill\quad\hfill 3 \hfill\quad\hfill 8 \hfill\quad\hfill 7 \hfill\quad\hfill 6 \hfill\quad\hfill 10 \hfill\quad\hfill 3 \hfill\quad\hfill 4 \hfill\quad\hfill 9$ & $1 \quad 2 \quad 5$ & $\hat{\psi}^{(5)}$ \\
$1 \hfill\quad\hfill 8 \hfill\quad\hfill 7 \hfill\quad\hfill 8 \hfill\quad\hfill 9 \hfill\quad\hfill 10 \hfill\quad\hfill 6 \hfill\quad\hfill 7 \hfill\quad\hfill 5 \hfill\quad\hfill 7$ & $2 \quad 3 \quad \boxed{4}$ & $\boxed{\hat{\psi}^{(6)}}$ \\
$2 \hfill\quad\hfill 6 \hfill\quad\hfill 5 \hfill\quad\hfill 2 \hfill\quad\hfill 1 \hfill\quad\hfill 4 \hfill\quad\hfill 5 \hfill\quad\hfill 8 \hfill\quad\hfill 6 \hfill\quad\hfill 10$ & $3 \quad 7 \quad 9$ & $\hat{\psi}^{(7)}$ \\
$10 \hfill\quad\hfill 10 \hfill\quad\hfill 2 \hfill\quad\hfill 5 \hfill\quad\hfill 5 \hfill\quad\hfill 2 \hfill\quad\hfill 3 \hfill\quad\hfill 6 \hfill\quad\hfill 10 \hfill\quad\hfill 9$ & $1 \quad \boxed{4} \quad 7 \quad 8$ & $\boxed{\hat{\psi}^{(8)}}$ \\
$4 \hfill\quad\hfill 8 \hfill\quad\hfill 6 \hfill\quad\hfill 1 \hfill\quad\hfill 6 \hfill\quad\hfill 6 \hfill\quad\hfill 1 \hfill\quad\hfill 5 \hfill\quad\hfill 7 \hfill\quad\hfill 6$ & $2 \quad 3 \quad 9 \quad 10$ & $\hat{\psi}^{(9)}$ \\
$8 \hfill\quad\hfill 4 \hfill\quad\hfill 6 \hfill\quad\hfill 7 \hfill\quad\hfill 7 \hfill\quad\hfill 7 \hfill\quad\hfill 10 \hfill\quad\hfill 10 \hfill\quad\hfill 5 \hfill\quad\hfill 10$ & $1 \quad 2 \quad 3 \quad 9$ & $\hat{\psi}^{(10)}$ \\
$1 \hfill\quad\hfill 8 \hfill\quad\hfill 6 \hfill\quad\hfill 5 \hfill\quad\hfill 5 \hfill\quad\hfill 2 \hfill\quad\hfill 1 \hfill\quad\hfill 8 \hfill\quad\hfill 6 \hfill\quad\hfill 6$ & $3 \quad \boxed{4} \quad 7 \quad 9 \quad 10$ & $\boxed{\hat{\psi}^{(11)}}$ \\
$4 \hfill\quad\hfill 6 \hfill\quad\hfill 10 \hfill\quad\hfill 5 \hfill\quad\hfill 10 \hfill\quad\hfill 9 \hfill\quad\hfill 10 \hfill\quad\hfill 9 \hfill\quad\hfill 9 \hfill\quad\hfill 9$ & $1 \hfill\quad\hfill 2 \hfill\quad\hfill 3 \hfill\quad\hfill 7 \hfill\quad\hfill 8$ & $\hat{\psi}^{(12)}$ \\
$9 \hfill\quad\hfill 2 \hfill\quad\hfill 8 \hfill\quad\hfill 4 \hfill\quad\hfill 10 \hfill\quad\hfill 1 \hfill\quad\hfill 1 \hfill\quad\hfill 9 \hfill\quad\hfill 6 \hfill\quad\hfill 3$ & $5 \quad 7$ & $\hat{\psi}^{(13)}$ \\
$2 \hfill\quad\hfill 2 \hfill\quad\hfill 8 \hfill\quad\hfill 9 \hfill\quad\hfill 1 \hfill\quad\hfill 10 \hfill\quad\hfill 2 \hfill\quad\hfill 9 \hfill\quad\hfill 5 \hfill\quad\hfill 10$ & $3 \quad \boxed{4} \quad 6 \quad 7$ & $\boxed{\hat{\psi}^{(14)}}$ \\
$4 \hfill\quad\hfill 4 \hfill\quad\hfill 1 \hfill\quad\hfill 4 \hfill\quad\hfill 1 \hfill\quad\hfill 8 \hfill\quad\hfill 4 \hfill\quad\hfill 3 \hfill\quad\hfill 1 \hfill\quad\hfill 4$ & $2 \quad 5 \quad 6 \quad 7 \quad 9 \quad 10$ & $\hat{\psi}^{(15)}$ \\
$10 \hfill\quad\hfill 2 \hfill\quad\hfill 1 \hfill\quad\hfill 7 \hfill\quad\hfill 9 \hfill\quad\hfill 8 \hfill\quad\hfill 4 \hfill\quad\hfill 2 \hfill\quad\hfill 2 \hfill\quad\hfill 10$ & $3 \quad 5 \quad 6$ & $\hat{\psi}^{(16)}$ \\
$2 \hfill\quad\hfill 8 \hfill\quad\hfill 7 \hfill\quad\hfill 10 \hfill\quad\hfill 9 \hfill\quad\hfill 2 \hfill\quad\hfill 1 \hfill\quad\hfill 5 \hfill\quad\hfill 7 \hfill\quad\hfill 6$ & $3 \quad \boxed{4}$ & $\boxed{\hat{\psi}^{(17)}}$ \\
$3 \hfill\quad\hfill 2 \hfill\quad\hfill 4 \hfill\quad\hfill 2 \hfill\quad\hfill 3 \hfill\quad\hfill 9 \hfill\quad\hfill 9 \hfill\quad\hfill 9 \hfill\quad\hfill 2 \hfill\quad\hfill 9$ & $1 \quad 5 \quad 6 \quad 7 \quad 8 \quad 10$ & $\hat{\psi}^{(18)}$ \\
$9 \hfill\quad\hfill 3 \hfill\quad\hfill 10 \hfill\quad\hfill 5 \hfill\quad\hfill 1 \hfill\quad\hfill 2 \hfill\quad\hfill 1 \hfill\quad\hfill 4 \hfill\quad\hfill 10 \hfill\quad\hfill 6$ & $7 \quad 8$ & $\hat{\psi}^{(19)}$ \\
$9 \hfill\quad\hfill 8 \hfill\quad\hfill 9 \hfill\quad\hfill 6 \hfill\quad\hfill 6 \hfill\quad\hfill 2 \hfill\quad\hfill 1 \hfill\quad\hfill 9 \hfill\quad\hfill 10 \hfill\quad\hfill 1$ & $3 \quad \boxed{4} \quad 5 \quad 7$ & $\boxed{\hat{\psi}^{(20)}}$ \\
$4 \hfill\quad\hfill 7 \hfill\quad\hfill 4 \hfill\quad\hfill 3 \hfill\quad\hfill 8 \hfill\quad\hfill 10 \hfill\quad\hfill 4 \hfill\quad\hfill 6 \hfill\quad\hfill 4 \hfill\quad\hfill 5$ & $1 \quad 2 \quad 9$ & $\hat{\psi}^{(21)}$ \\
$5 \hfill\quad\hfill 1 \hfill\quad\hfill 9 \hfill\quad\hfill 6 \hfill\quad\hfill 5 \hfill\quad\hfill 9 \hfill\quad\hfill 1 \hfill\quad\hfill 8 \hfill\quad\hfill 4 \hfill\quad\hfill 5$ & $2 \quad 3 \quad 7 \quad 10$ & $\hat{\psi}^{(22)}$ \\
$5 \hfill\quad\hfill 5 \hfill\quad\hfill 3 \hfill\quad\hfill 1 \hfill\quad\hfill 1 \hfill\quad\hfill 6 \hfill\quad\hfill 1 \hfill\quad\hfill 1 \hfill\quad\hfill 7 \hfill\quad\hfill 6$ & $2 \quad \boxed{4} \quad 8 \quad 9 \quad 10$ & $\boxed{\hat{\psi}^{(23)}}$ \\
$4 \hfill\quad\hfill 10 \hfill\quad\hfill 5 \hfill\quad\hfill 9 \hfill\quad\hfill 6 \hfill\quad\hfill 5 \hfill\quad\hfill 1 \hfill\quad\hfill 6 \hfill\quad\hfill 1 \hfill\quad\hfill 9$ & $2 \quad 3 \quad 7 \quad 8$ & $\hat{\psi}^{(24)}$ \\
$5 \hfill\quad\hfill 5 \hfill\quad\hfill 5 \hfill\quad\hfill 9 \hfill\quad\hfill 8 \hfill\quad\hfill 2 \hfill\quad\hfill 4 \hfill\quad\hfill 9 \hfill\quad\hfill 1 \hfill\quad\hfill 5$ & $3 \quad 6 \quad 7 \quad 10$ & $\hat{\psi}^{(25)}$ \\
\hline\hline
\end{tabular}
\label{tab:oob}
\end{table}}

\subsection{Two-stage out-of-bag conformal prediction}\label{ssec:tsoobcp}

In this section, we restrict our attention to using only random forests in the frequency and the severity modeling stages, guiding ourselves by the general idea that calibration sample quantities in the two-stage split conformal prediction Algorithm \ref{algo:tsscp} can be represented by the corresponding out-of-bag predictions made within the training sample, thereby avoiding the data splitting process involved in Algorithm \ref{algo:tsscp}.

We have again the random sample $(X_1,D_1,Y_1),\dots,(X_n,D_n,Y_n),(X_{n+1},D_{n+1},Y_{n+1})$ described in Section \ref{sec:freqsev}. Unlike our approach in Section \ref{ssec:twostage}, there is no data splitting: the first $n$ triplets constitute our training sample.

For the frequency stage, we use the whole training sample to construct a random forest $\hat{\mu}=\{\hat{\mu}^{(j)}\}_{j\in B}$, with $B$ trees. For $i=1,\dots,n$, we determine the random sub-forest $\{\hat{\mu}^{(j)}:j\in\mathcal{O}_i\}$ for which the $i$-th training sample unit stayed out-of-bag of the corresponding bootstrap samples, collecting in $\mathcal{O}_i\subset\{1,\dots,B\}$ the respective indices of these regression trees. Using this random sub-forest we compute the out-of-bag predictions for the claim frequency $\hat{d}_i=|\mathcal{O}_i|^{-1} \sum_{j\in\mathcal{O}_i}\hat{\mu}^{(j)}(x_i)$, for $i=1,\dots,n$.

Moving to the severity stage, we train a random forest $\hat{\psi}=\{\hat{\psi}^{(j)}\}_{j\in B'}$, with $B'$ trees, from $\{(x_i,\hat{d}_i,y_i)\}_{i=1}^n$. This introduction of the predicted claim frequency is necessary to create a symmetric treatment of the conformity scores computed with the training sample units and a future triplet $(X_{n+1},D_{n+1},Y_{n+1})$, for which the only observed input available to compute the prediction interval for $Y_{n+1}$ will be the vector of predictors $X_{n+1}$. Similarly to what we have done for the frequency random forest $\hat{\mu}$, we keep the out-of-bag information for the severity random forest $\hat{\psi}$ in $\mathcal{O'}_i\subset\{1,\dots,B'\}$.

Finally, we compute the out-of-bag residuals $\delta_i=\left|y_i - |\mathcal{O'}_i|^{-1} \sum_{j\in\mathcal{O'}_i}\hat{\psi}^{(j)}(x_i)\right|$, for $i=1,\dots,n$, and train a severity variability random forest $\hat{\sigma}=\{\hat{\sigma}^{(j)}\}_{j\in B''}$ with $B''$ trees, from the sample $\{(x_i,\hat{d}_i,\delta_i)\}_{i=1}^n$. Out-of-bag information for $\hat{\sigma}$ is kept in $\mathcal{O''}_i\subset\{1,\dots,B''\}$, similarly to what we have done before for $\hat{\mu}$ and $\hat{\psi}$. Furthermore, conformity scores are now computed within the training sample as
$R_i = \delta_i \big/ \left( |\mathcal{O''}_i|^{-1} \sum_{j\in\mathcal{O''}_i}\hat{\sigma}^{(j)}(x_i,\hat{d}_i) \right)$, for $i=1,\dots,n$, and the conformal prediction interval for a future severity $Y_{n+1}$ is given by
$$
  C^{(1-\alpha)}(X_{n+1}) = [\,\max\,\{0, \hat{\psi}(X_{n+1},\hat{\mu}(X_{n+1})) - \epsilon\}, \hat{\psi}(X_{n+1},\hat{\mu}(X_{n+1})) + \epsilon\,],
$$
in which $X_{n+1}$ is a future vector of predictors and $\epsilon = R_{(\lceil(1-\alpha)(n+1)\rceil)} \times \hat{\sigma}(X_{n+1},\hat{\mu}(X_{n+1}))$. A programmatic description of this two-stage out-of-bag conformal prediction procedure is given in Algorithm \ref{algo:tsoobcp}. In the next section, we discuss the performance of Algorithm \ref{algo:tsoobcp} when applied to the datasets analyzed in Section \ref{sec:examples}.

\begin{algorithm}[t!]
\caption{Two-stage out-of-bag conformal prediction}\label{algo:tsoobcp}
\begin{algorithmic}[1]
  \Require Dataset $\{(x_i,d_i,y_i)\}_{i=1}^n$, numbers $B,B',B''$ of trees used to train the frequency random forest, the severity random forest, and the severity variability random forest, respectively, future vector of predictors $x_{n+1}\in\mathbb{R}^p$, and nominal miscoverage level $0<\alpha<1$.
  \Ensure Prediction interval.
  \Statex
  \Function{oob}{sample unit index $i$, random forest $\hat{\tau}=\{\hat{\tau}^{(j)}\}_{j=1}^B$}
    \State $\mathcal{O}\gets\emptyset$
    \For{$j \gets 1 \textrm{ to } B$}
      \If{$i$-th sample unit is not included in the $j$-th bootstrap sample}
        \State $\mathcal{O} \gets \mathcal{O} \cup \{j\}$
      \EndIf
    \EndFor
    \State\Return $\mathcal{O}$
  \EndFunction
  \Statex
  \State Train frequency random forest $\hat{\mu}=\{\hat{\mu}^{(j)}\}_{j=1}^B$ from $\{(x_i,d_i)\}_{i=1}^n$
  \For{$i=1 \textrm{ to } n$}
    \State $\mathcal{O}_i \gets$ \textsc{oob}($i,\hat{\mu}$)
    \State $\hat{d}_i \gets \frac{1}{|\mathcal{O}_i|} \sum_{j\in\mathcal{O}_i}\hat{\mu}^{(j)}(x_i)$
  \EndFor
  \State Train severity random forest $\hat{\psi}=\{\hat{\psi}^{(j)}\}_{j=1}^{B'}$ from $\{(x_i,\hat{d}_i,y_i)\}_{i=1}^n$
  \For{$i=1 \textrm{ to } n$}
    \State $\mathcal{O'}_i \gets$ \textsc{oob}($i,\hat{\psi}$)
    \State $\delta_i \gets \left|y_i - \frac{1}{|\mathcal{O'}_i|} \sum_{j\in\mathcal{O'}_i}\hat{\psi}^{(j)}(x_i)\right|$
  \EndFor
  \State Train severity variability forest $\hat{\sigma}=\{\hat{\sigma}^{(j)}\}_{j=1}^{B''}$ from $\{(x_i,\hat{d}_i,\delta_i)\}_{i=1}^n$
  \For{$i=1 \textrm{ to } n$}
    \State $\mathcal{O''}_i \gets$ \textsc{oob}($i,\hat{\sigma}$)
    \State $r_i \gets \delta_i \Big/ \left( \frac{1}{|\mathcal{O''}_i|} \sum_{j\in\mathcal{O''}_i}\hat{\sigma}^{(j)}(x_i,\hat{d}_i) \right)$
  \EndFor
  \State $\epsilon \gets r_{(\lceil(1-\alpha)(n+1)\rceil)} \times \hat{\sigma}(x_{n+1},\hat{\mu}(x_{n+1}))$
  \State \Return{$[\,\max\,\{0, \hat{\psi}(x_{n+1},\hat{\mu}(x_{n+1})) - \epsilon\}, \hat{\psi}(x_{n+1},\hat{\mu}(x_{n+1})) + \epsilon\,]$}
\end{algorithmic}
\end{algorithm}

\subsection{Out-of-bag performance}\label{ssec:oobperf}

For the three datasets discussed in Section \ref{sec:examples}, the use of Algorithm \ref{algo:tsoobcp} to produce the conformal prediction intervals allows us to enlarge our training samples, since it operates without the use of the calibration sample needed for Algorithm \ref{algo:tsscp}. Through all the examples in this section, we use random forests with 1,000 trees in Algorithm \ref{algo:tsoobcp}. For the sake of comparison, we systematically make the new training sample to be the union of the original training and calibration samples used in the data splitting methods of each example in Section \ref{sec:examples}. Consequently, the resulting test samples are the same used in the examples discussed in Section \ref{sec:examples}, allowing the direct comparison of the corresponding test performances.

\begin{figure}[t!]
\centering
\includegraphics[width=16cm]{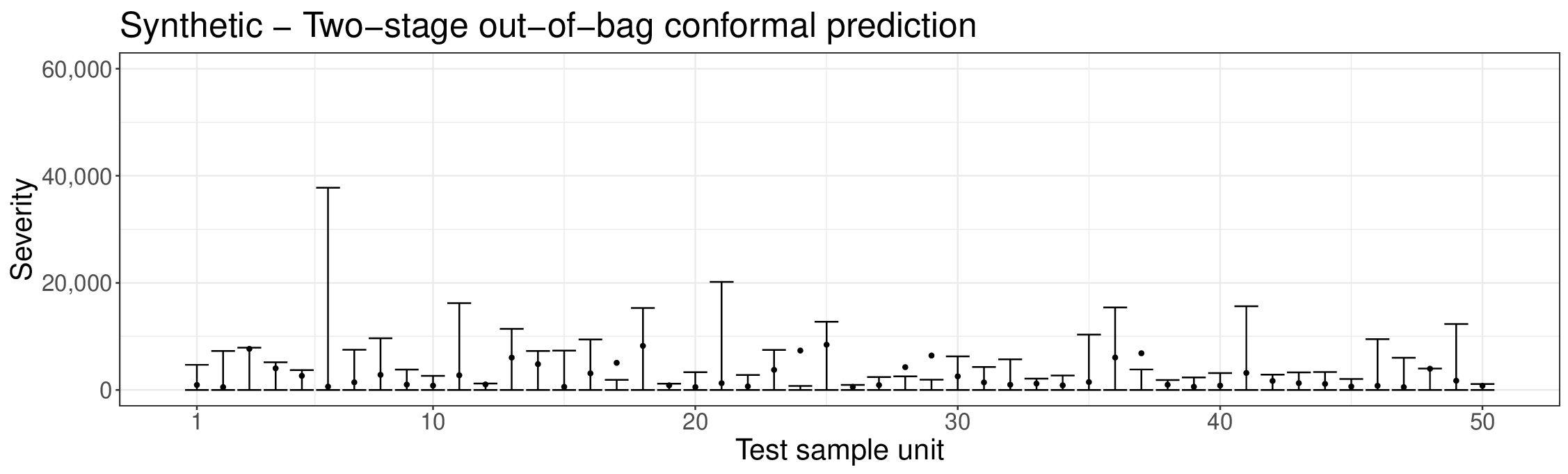}
\caption{Two-stage out-of-bag conformal prediction sets produced by Algorithm \ref{algo:tsoobcp} for the same fifty test sample units of the synthetic dataset appearing in Figure \ref{fig:synthsplit}.}
\label{fig:synthoob}
\end{figure}

Conformal prediction sets generated with Algorithm \ref{algo:tsoobcp} for the synthetic dataset have a test sample coverage of 91.11\%. An efficiency gain is also observed, as we have an average prediction interval width equal to 8,659.70, which is a 28.80\% decrease with respect to the best data splitting method (12,161.69) presented in Section \ref{ssec:synth}. Figure \ref{fig:synthoob} shows the prediction intervals obtained with Algorithm \ref{algo:tsoobcp} for the same fifty test sample units in the synthetic dataset used in Figure \ref{fig:synthsplit}.

Moving to the analysis of the MTPL dataset discussed in Section \ref{ssec:mtpl}, conformal prediction sets produced by Algorithm \ref{algo:tsoobcp} for this example have a test sample coverage of 91.34\%. We observe an even better efficiency gain in the prediction intervals produced for the test sample in the MTPL dataset, which have an average width equal to 398,28, representing a remarkable 70.15\% decrease with respect to the best data splitting method (1,334.31) presented in Section \ref{ssec:mtpl}. For comparison, Figure \ref{fig:synthoob} shows the prediction intervals obtained with Algorithm \ref{algo:tsoobcp} for the same fifty test sample units in the MTPL dataset displayed in Figure \ref{fig:mtplsplit}.

Finally, for the crop insurance dataset analyzed in Section \ref{ssec:crop}, conformal prediction sets produced by Algorithm \ref{algo:tsoobcp} for the crop dataset have a test sample coverage of 90.67\%. A substantial efficiency gain is also observed when using Algorithm \ref{algo:tsoobcp} for the crop insurance dataset: the prediction intervals produced for the test sample have an average width equal to 158,655.10, which is a 54.54\% decrease with respect to the best data splitting method (349,026.40) presented in Section \ref{ssec:crop}. For comparison, Figure \ref{fig:synthoob} shows the prediction intervals obtained with Algorithm \ref{algo:tsoobcp} for the same fifty test sample units in the crop insurance dataset presented in Figure \ref{fig:mtplsplit}. All the results for the three datasets using the different methods are summarized in Table \ref{tab:comparison}.

\begin{figure}[t!]
\centering
\includegraphics[width=16cm]{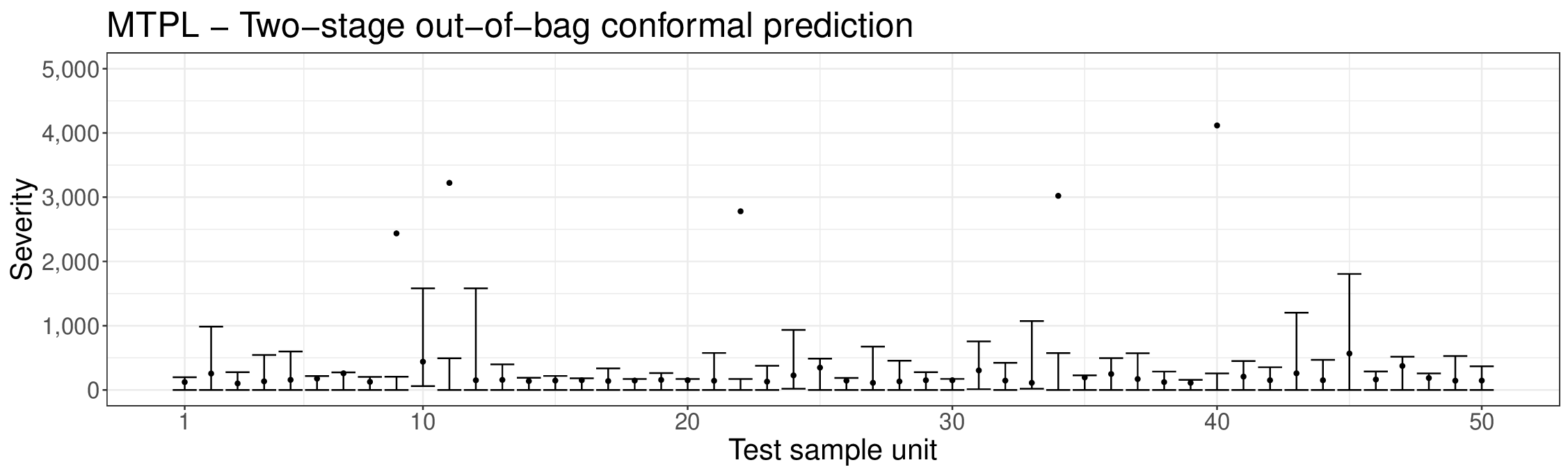}
\caption{Two-stage out-of-bag conformal prediction sets produced by Algorithm \ref{algo:tsoobcp} for the same fifty test sample units in the motor third party liability dataset appearing in Figure \ref{fig:mtplsplit}.}
\label{fig:mtploob}
\end{figure}

\begin{figure}[t!]
\centering
\includegraphics[width=16cm]{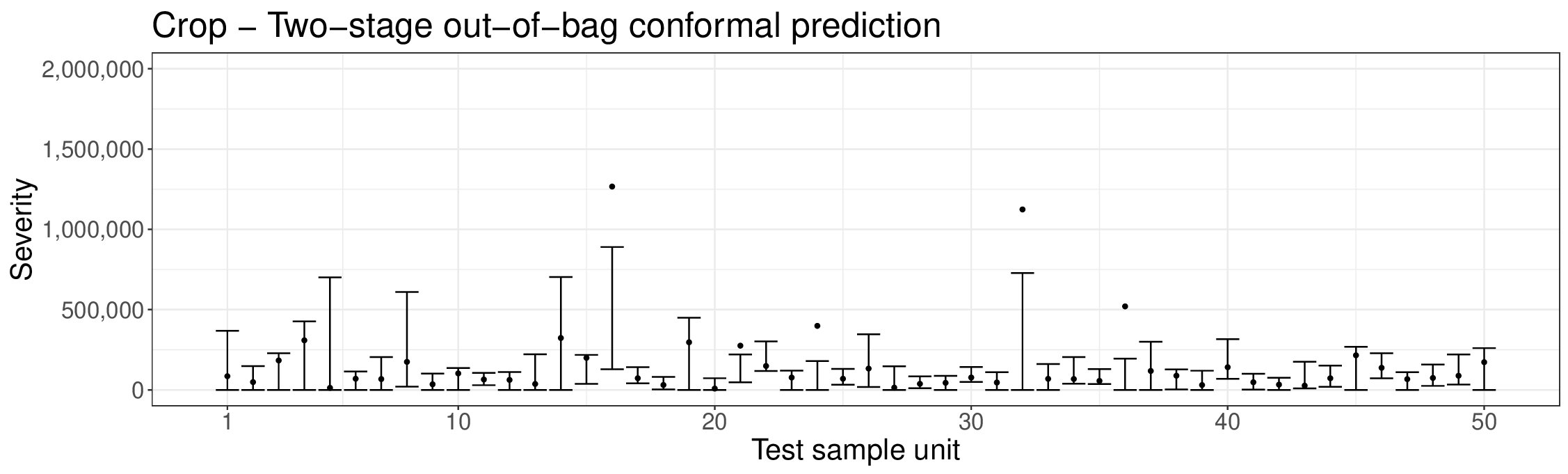}
\caption{Two-stage out-of-bag conformal prediction sets produced by Algorithm \ref{algo:tsoobcp} for the same fifty test sample units in the crop dataset appearing in Figure \ref{fig:cropsplit}.}
\label{fig:cropoob}
\end{figure}

\section{Concluding remarks}\label{sec:concl}

The increasing use of machine learning methods for predictive modeling in the insurance industry creates the need for a systematic assessment of the uncertainty in the forecasts made by the plethora of available algorithms. Ideally, this framework should not rely on strong assumptions about the underlying data generating process and should contain statistical guarantees regarding the coverage of the prediction intervals produced by the method. This paper contributes in this direction, proposing conformal prediction procedures suitable for dealing with the two-stage nature of the frequency-severity modeling process used in actuarial problems. A natural alternative to the conformal methods presented in this study would be to use classical parametric bootstrap techniques in the setup where a gamma regression is used to model the severity stage. Although bootstrap methods have been successfully used in different actuarial applications -- such as the chain ladder model \cite{steinmetz2022, steinmetz2024} commonly used to build up a reserve to meet future obligations arising from incurred claims -- in our frequency-severity modeling setting the parametric bootstrap \cite{haman2022,gelman2007} performed poorly both in terms of coverage and average width of the prediction intervals produced by the method. A final comparison of all the methods considered in the paper is given in Table \ref{tab:comparison}, in which we see, for each considered dataset, that severity models with better generalization capacity yield prediction intervals with shorter average width.

The procedures discussed in the paper can be generalized to multi-stage settings. For example, in the case of the Crop insurance dataset discussed in Section \ref{ssec:crop}, we could have a first-stage model predicting the weather features, followed by frequency and severity models at the second and third stages, respectively. Finally, in future work, an extension of the proposed methods to non-exchangeable cases could be effected by exploring a break up in the distributional symmetry of the data like the one discussed in \cite{tibshirani2019}.

\begin{table}[t!]
\small
\centering
\captionof{table}{Performance comparison among the different models and methods applied to the three datasets.}\label{tab:comparison}
\begin{tabular}{lllcr}
\hline\hline
Dataset & Model & Method & Coverage & Average interval width \\
\hline\hline
Synthetic & Gamma & Split & 90.25\% & 23,717.68  \\
Synthetic & Random Forest & Split & 90.29\% & 12,161.26 \\
Synthetic & Random Forest & OOB & 91.11\% & 8,659.70 \\
Synthetic & Gamma & Bootstrap & 30.83\% & 50,789.91 \\
\hline
MTPL & Gamma & Split & 89.70\% & 2,902.11 \\
MTPL & Random Forest & Split & 89.93\% & 1,334.31 \\
MTPL & Random Forest & OOB & 91.34\% & 398.28 \\
MTPL & Gamma & Bootstrap & 11.05\% & 7,656.35 \\
\hline
Brazilian crop & Gamma & Split & 90.71\% & 459,556.80 \\
Brazilian crop & Random Forest & Split & 90.50\% & 349,026.40 \\
Brazilian crop & Random Forest & OOB & 90.67\% & 158,655.10 \\
Brazilian crop & Gamma & Bootstrap & 29.62\% & 616,466.00 \\
\hline\hline
\end{tabular}
\end{table}

\section*{Appendix: Open source software and data availability}\label{sec:appendix}

Open source software, coded in \texttt{R} \cite{R}, and data for all the examples in the paper are available at \url{https://github.com/heltongraziadei/conformal-fs}. In this repository, we have three folders, named \texttt{synthetic}, \texttt{mtpl}, and \texttt{crop}, corresponding to the analyses of the respective datasets discussed in Section \ref{sec:examples}. Inside each folder, a suffix on a script name identifies the method used to produce the prediction intervals. Suffixes \texttt{split\_gamma} and \texttt{split\_rf} refer to applications of the data splitting Algorithm \ref{algo:tsscp} using gamma regressions and random forests as models for the severity stage, respectively. The suffix \texttt{oob} refers to applications of Algorithm \ref{algo:tsoobcp}. Finally, the suffix \texttt{bootstrap} identifies code for the bootstrap procedure mentioned in the concluding remarks, Section \ref{sec:concl}.

\section*{Acknowledgements}

Paulo C. Marques F. receives support from FAPESP (Fundação de Amparo à Pesquisa do Estado de São Paulo) through project 2023/02538-0. Rodrigo S. Targino acknowledges support from CNPq (200293/2022-2) and FAPERJ (E-26/201.350, E-26/211.426, E-26/211.578).

\bibliographystyle{ieeetr}
\bibliography{bibliography}

\end{document}